\documentstyle[aps,prb]{revtex}
\begin{document}
\twocolumn
\draft

\title{Observation and Assignment of Silent and Higher Order Vibrations in
the Infrared Transmission of C$_{60}$ Crystals}

\author{Michael C. Martin, Xiaoqun Du, John Kwon, and L. Mihaly}

\address{Department of Physics, State University of New York at Stony Brook,
Stony Brook, NY 11794--3800}

\date{Submitted to Phys. Rev. B, January 24, 1994}
\maketitle

\begin{abstract}
We report the measurement of infrared transmission of large C$_{60}$ single
crystals.  The spectra exhibit a very rich structure with over 180
vibrational absorptions visible in the $100 - 4000 $cm$^{-1}$ range.  Many
silent modes are observed to have become weakly IR-active.
We also observe a large number of higher order combination modes.
The temperature (77K$ - 300$K) and pressure ($0 - 25$KBar) dependencies of
these modes were measured and are presented.  Careful analysis of the IR
spectra in conjunction with Raman scattering data showing second order modes
and neutron scattering data, allow the selection of the 46 vibrational modes
C$_{60}$.  We are able to fit {\it all} of the first and second order data
seen in the present IR spectra and the previously published Raman data
($\sim 300$ lines total), using these 46 modes and their group theory
allowed second order combinations.
\end{abstract}
\pacs{PACS: 78.30.-j, 63.20.-e, 63.20.Ry}

\narrowtext
\section{Introduction}

Buckminsterfullerene has become one of the most studied materials in recent
years.  Its very symmetric structure, and the discovery of superconductivity
in doped C$_{60}$ \cite{discsuper}, has spurred a great interest in the
vibrational modes of the C$_{60}$ molecule.  Many theoretical models
\cite{Wu,Stanton,Negri,Weeks,Adams,Millie,quong} have been presented to
predict properties of the 46 distinct modes allowed by group theory.

The truncated icosohedral structure of C$_{60}$ fullerenes belongs to the
icosohedral point group, $I_{h}$, and has 4 infrared active intramolecular
vibrational modes with F$_{1u}$ symmetry.  There are also 10 Raman-active
vibrational modes:  2 with A$_g$ symmetry and 8 with H$_g$ symmetry.  All 14
of these modes have been experimentally observed with great precision by
many researchers \cite{earlyIR,Kratschmer,raman,raman1}.  32 additional
modes that are IR  and Raman forbidden (silent modes) are 1A$_u$, 3F$_{1g}$,
4F$_{2g}$, 5F$_{2u}$, 6G$_{g}$, 6G$_{u}$, and 7H$_{u}$.  The optical modes,
derived from these silent vibrations can in
principle be measured directly by inelastic neutron scattering and electron
scattering experiments, however these measurements
\cite{neutron1,eels} lack the high resolution obtainable by optical probes.

In this work we present infrared transmission data obtained on C$_{60}$
crystals thick enough to quantitatively measure over 180 weak vibrational
modes.  Temperature and pressure dependencies of these modes are shown.  32
fundamental frequencies of the silent vibrational modes were then extracted
by careful analysis of a large number of weakly IR-active features observed
in these spectra in conjunction with previous Raman \cite{eklundraman} and
Neutron \cite{neutron1} measurements.  Using group theoretical
considerations, we are able to fit all of the modes seen weakly in IR and
Raman spectra.

There are several ways for weak IR lines to appear in a C$_{60}$ sample.
Some of these work for a single molecule (although the signal would be
vanish in the noise for a gas phase sample), others require the presence of
the interactions typical of a solid.  First, any vibrational mode may become
weakly IR-active due to an isotopic impurity in some C$_{60}$ molecules
\cite{eklundraman}.
The natural abundance of $^{13}$C is approximately 1.1\%, therefore one
would expect that 34.4\% of the fullerene molecules have one $^{13}$C atom,
({\it i.e.} the molecule would be $^{13}$C$^{12}$C$_{59}$).  The addition
of a single $^{13}$C lowers the symmetry of the molecule and is expected
to make {\it all} of the previously silent modes active.  This isotopic
activation of vibrations has been similarly observed in benzene
\cite{benzene}.

Weak IR activity may appear due to the molecule being located at an FCC
symmetry site in the solid.  The crystal fields reduces the icosohedral
symmetry of the fullerenes and activates normally silent odd-parity modes
\cite{vanloos}.  Symmetry breaking from electric field gradients due to
surface effects, impurities and dislocations, also provide ways for silent
modes to appear in the IR spectrum.

The anharmonicity of the bonding potential leads to new lines, called
combination modes, appearing at frequencies different from the fundamental
resonances.  Typically the anharmonicity can be treated perturbatively, and
the calculation leads to weak modes at frequencies close to the sums and
differences of the fundamental vibrational frequencies.  If two fundamentals
are involved, a ``binary" combination is produced.  The ``overtone" is a
special binary combination, appearing at twice the fundamental frequency.
The symmetry of the molecule constrains the possible IR or Raman active
combination modes.

In principle two phonon processes ({\it i.e.} the emission of a Raman-active
phonon followed by the excitation of an IR-active mode) can lead to the
appearance of weak IR features at the sum and difference frequencies
involving Raman-active and IR-active mode.  Since the Raman cross section
drops dramatically at lower incident photon energies, this process is
expected to be negligible relative to the other processes discussed above.

Several authors reported observation of weak modes in Raman
\cite{vanloos,eklundraman} or IR
\cite{irmodes,kamaras1,kamaras2,eklundhigherorderir} spectroscopy.
In the first report containing a
careful and detailed analysis of combination modes in Raman data, Z.-H. Dong
{\it et al.} \cite{eklundraman} reported higher order Raman lines up to
3000 cm$^{-1}$.  Their fundamental frequencies were chosen to be close to a
force-constant model \cite{Millie} prediction, slightly modified to
better fit the Raman data.  All the allowed combinations of these
theoretically predicted modes were calculated and compared the Raman data.
Kamar\'as {\it et al.} observed weak IR modes in C$_{60}$ films \cite{kamaras1}
and in C$_{60}$ n-pentane single crystals \cite{kamaras2}, and
assigned some of these to Raman or other IR-silent fundamental vibrational
modes.  The most detailed recent IR study has been done by K.-A. Wand
{\it et al.} \cite{eklundhigherorderir}.  They measured fairly thick
C$_{60}$ films on a KBr substrate and observe very weak features exactly
where we report vibrational modes in the present paper.  Their analysis
was similar to the Raman study \cite{eklundraman} with a refinement of
the theoretical fundamental mode frequencies.

The present study is similar to the one by Dong {\it et al.}
\cite{eklundraman} and Wand {\it et al.} \cite{eklundhigherorderir},
in that we are selecting fundamental frequencies to fit the second order
combination modes.  However, our crystals are much thicker than the
films used by Wand {\it et al.},  making the vibrational modes much
clearer and showing even more modes weaker than could be observed in their
study.  Whenever comparison was possible, there was no visible disagreement
in the experimental data, but we found that the fundamentals chosen in the
previous works did not give a satisfactory fit to the extended data set.
Furthermore, by looking at the temperature dependence of the spectra,
the difference modes can be excluded from consideration (difference
modes were used by Dong {\it et al.} to fit the Raman measurements).
Therefore we were forced to take an approach where our analysis has
not been based on any theoretical model of C$_{60}$ vibrations.
Instead as much information as possible was extracted from the
experimental data itself.  The
values obtained are in a few cases reasonably close to the
values used by Wand {\it et al.}, however most deviate enough to
change the assignments of the higher order modes significantly. In the
process we also learned that even with the considerable constrains provided
by the combination of our experimental data, the Raman data of
Dong {\it et al.}, and neutron scattering data \cite{neutron1},
the assignment of the fundamentals is still not unique.

The organization of this paper is as follows.  First, the samples and the
experimental procedures are described.  Next the results at various
temperatures are discussed, then we turn to the pressure dependence.
Subsequently, the principles used in the the analysis of the experimental
data are summarized and our results are presented.  Finally, our results
are discussed in the context of the neutron scattering and other
spectroscopic studies.

\section{Experimental}

The C$_{60}$ for this study was purchased from SES Research, Inc. \cite{ses}.
The fullerene powder
was placed in a quartz tube, baked out under vacuum for 1 day at $250^{o}$C
to remove volatile contaminants, and sealed under a low pressure of Argon
gas.  This tube was then placed in a specially made oven where the C$_{60}$
powder was kept at $630^{o}$C and the opposite end of the sealed tube
maintained at $450^{o}$C.  C$_{60}$ crystals grew by vapor transport over
the course of approximately 2 weeks. The crystals are as large as $\sim
1\times 1\times 0.5 $mm$^{3}$.  The crystals were mounted free standing over
an aperture approximately 0.5 mm in diameter with a minimal amount of silver
paste to hold them in place.

For the study of the temperature dependence, the samples were then mounted
on the cold finger of a Heli-Tran open cycle cryostat.  Measurements were
performed between room temperature and 77K.  The pressure dependence was
studied in a High Pressure Diamond Optics, Inc. diamond anvil cell at room
temperature.  First the crystal was inserted in the 0.3 mm diameter hole of
the stainless steel gasket and placed between the diamonds.  No pressure
transmitting medium was applied.  At this point the sample did not fill the
gasket completely, but as the diamonds were compressed the crystal broke,
and very soon the gasket was entirely filled with the C$_{60}$.
Transmission measurements were feasible after this point, although in these
measurements the sample was not a single crystal, but rather a collection of
crystallites.  The pressure was calibrated by using the known pressure
dependence of the F$_{1u}$ IR-active modes \cite{c60pres}.  The data
discussed in this paper was obtained by first increasing the pressure to
$\sim25$KBar and then recording spectra as the pressure was slowly reduced.
The last of these spectra was obtained at ambient pressure.

Some of the infrared measurements were carried out at the National
Sychrotron Light Source, Brookhaven National Laboratory.  The far-infrared
spectra ($100-600 $cm$^{-1}$ region) were obtained with a Nicolet 20F Rapid
Scan FTIR spectrometer at beamline U4IR at a resolution of 1.5cm$^{-1}$.
The mid-infrared spectra ($500-4000 $cm$^{-1}$) were obtained using a Nicolet
740 FTIR spectrometer on beamline U2B at a resolution of 0.5cm$^{-1}$.
This latter region was also obtained at Stony Brook using a Bomem MB-155
FTIR spectrometer and a standard globar source and was found to be
identical.  Apart from variations due to different sample thicknesses, the
spectra varied extremely little with different C$_{60}$ crystals grown from
different batches.

Throughout this study weak features in the optical response of the sample
are being investigating.   A simple computer simulation of an ``ideal"
measurement illustrates the usefulness of the transmission method as
compared to reflectivity measurements for very weak modes.  Figure
\ref{transrefl} presents the calculated response of two oscillators in
transmission and reflectivity.  The reflectivity was calculated from ${\sl
R}=((1-n)^2+k^2)/(1+n)^2+k^2)$, where n and k are the real and imaginary
part of $\sqrt {\epsilon (\omega )}$.  This formula assumes infinite sample
thickness.  The transmission is given by $\vert t \vert ^2$, where
\begin{equation}
t=\frac{4(n+ik)}{(1+n+ik)e^{-i2\pi d(n+ik)\omega }-(1-n-ik)
e^{i2\pi d(n+ik)\omega }}
\label{transmission}
\end{equation}
The sample thickness $d$ was chosen to be $d=0.5$mm for Figure
\ref{transrefl}.  The dielectric function is described by $\epsilon (\omega
)=\sum _i \epsilon _i + \epsilon _\infty $, where the i-th oscillator is
represented by
\begin{equation}
\epsilon _i = \frac{\omega _p^2}{\omega _0^2-\omega ^2-i\Gamma \omega }
\label{oscillator}
\end{equation}
where $\omega _p$ is the plasma frequency, $\omega _0$ is the resonance
frequency, $\Gamma $ is the the width and $\epsilon _\infty $ is the
dielectric constant due to other oscillators located at higher frequencies.
(the strength, $S=\omega _p^2 / \omega _0^2$, characterizes the
contribution of each oscillator to the dielectric constant at lower
frequencies).  With the parameters indicated in the Figure, one of the
oscillators is ``strong", corresponding to an IR-active mode, while the
other one is about 400 times weaker, representing a ``silent" or higher
order mode.  The ``strong" mode is easily visible in transmission and
reflection as well, but the weak one shows up only in transmission.
(Closer inspection of the reflectivity curve reveals a
$\Delta R = 0.001$\% modulation at the weak resonance).

Figure \ref{transrefl} may also help to convert the
features seen in the raw data to intrinsic properties of the sample.  The
regular oscillations in transmission background are due to interference
between the the light reflected from the front and back surface of the
sample.  The period of these oscillations is simply related to the thickness
of the sample and the index of refraction by $\Delta (1/\lambda )= 1/(2nd)$.
Under real experimental conditions,
the oscillations can be easily smeared out by non-uniform thickness
of the sample, but in our single crystals they are quite visible.
Notice also that for strong oscillators the width of the
transmission minimum has very little relation to the true width of the
resonance, and the resonance frequency $\omega _0$ is {\it not} at the
middle of the transmission feature.

Eqns. \ref{transmission} and \ref{oscillator} suggest that an arbitrarily
weak oscillator may be seen in transmission measurements on samples of
sufficient thickness.  Of course, in real experimental situations this
is not the case.  The main experimental limitation comes form the
imperfect nature of the sample, leading to additional absorption and
scattering in the bulk of the specimen, in spite of Eqn. \ref{transmission}
suggesting that (apart from the oscillators) the transmission
background is independent of the sample thickness.  Consequently, very thick
samples have very small overall transmission.  This points to another aspect
of our present investigation: the use of good quality single crystals.  Thin
films, typically used in transmission studies, have defects (mostly
grain boundaries) leading to dramatically reduced transmission at modest
thickness.  With our single crystals we were indeed able to perform
transmission measurements on 0.5mm thick samples.  These crystals, like all
bulk C$_{60}$ samples, look dark-grey/black and are totally opaque in
visible light, but they still transmit at about 50\% of the ``ideal" value
in the IR regime (assuming, of course, that one looks at a frequency
sufficiently far from a resonance).

\subsection{C$_{60}$ IR Spectra and their Temperature Dependence}

We present the infrared transmission data on a C$_{60}$ single crystal at
300 and 77K in Figure \ref{IR&modes}.  There are approximately 150
vibrational mode absorptions visible in the room temperature spectrum and
many of the modes have a fine structure at low T.
The four F$_{1u}$ modes are seen to be so strong that they saturate
with zero transmission in a range around the usual 527, 576, 1182 and
1429cm$^{-1}$ positions.

We first looked for any sign of contaminants
in the crystals that could give rise to any of the additional weak modes.
The possibility that any solvents used in the C$_{60}$ purification remain
in the crystals can be excluded since they all have very strong, broad
absorptions near 2900cm$^{-1}$ that are not seen in our spectra.  Similarly,
our spectra is compared to published infrared results for C$_{70}$
\cite{c70} and C$_{60}$O \cite{c60o} and it is concluded that our crystals
do not contain either compound.

Many of the known Raman vibrational modes can readily be identified in our
infrared spectra using their well known positions:  H$_g(2)$ at 431cm$^{-
1}$, H$_g(3)$ at 709cm$^{-1}$, H$_g(4)$ at 775cm$^{-1}$, H$_g(5)$ at
1102cm$^{-1}$, H$_g(8)$ around 1576cm$^{-1}$, and A$_g(2)$ at 1470cm$^{-1}$.
H$_g(7)$ at 1425cm$^{-1}$ might also be active, but it is hidden under the
extremely strong F$_{1u}(4)$ mode nearby.  H$_g(1)$, H$_g(6)$ and A$_g(1)$
do not appear in our IR spectra.

The observation and identification of the A$_g(2)$ Raman line in the IR
transmission spectrum provides us with another way of characterizing the
sample.  As detailed Raman studies by Rao {\it et al.} \cite{rao} and
Kuzmany {\it et al.} \cite{kuzmanyox,kuzmany} have
demonstrated, in samples exposed to oxygen this mode shifts in frequency,
most likely due to the binding of the O to the C$_{60}$
molecules.  On the other hand, in oxygen free environment, the C$_{60}$
solid has a tendency to polymerize, facilitated by the exposure to visible
light (typically the laser used in Raman spectroscopy)
\cite{polymerization,kuzmany}.  The polymerized sample also has a
downshifted A$_g(2)$ line.  According to the Raman studies the oxygen
uptake or the polymerization is very fast.

Our samples were exposed to moderate amounts of visible light before and
after opening the crystal growth tubes and were held in air for several
days/weeks during the measurements.  Still, the inspection of the A$_g(2)$
line indicates that the bulk of our crystals is neither oxygenated, nor
polymerized.  To confirm this conclusion, a crystal treated in a
completely oxygen free environment was measured resulting in
identical spectra over the whole range of frequencies, including the
neighborhood of the A$_g(2)$ mode.  Furthermore, the oxygen free crystal
was exposed to intense UV light for a week and the partial shift of the
1470cm$^{-1}$ line to lower frequencies was observed indicating
polymerization had occurred (confirming that this line does correspond to
the A$_g(2)$ vibrational mode, and that the sample was indeed oxygen free).
We believe that the rapid oxygen uptake or polymerization is a property of
thin polycrystalline films, whereas the bulk of the single crystals is well
protected from oxygen (by the limited diffusion) and from light (by
the finite penetration depth of the short wavelength photons).  It is
also possible that the exposure to visible light
before the sample tube was opened created a protective polymer layer
on the surface of the crystals which does not allow oxygen to enter
the crystal bulk.

The appearance of the Raman lines in the IR spectra strongly suggests that
other IR-inactive resonances could appear in the spectra as well.  However,
the total number of lines far exceeds the number of vibrational modes, and
the extension of the spectral features up to $\sim 3200$cm$^{-1}$ indicates
that combination modes are also present.  The binary combinations (involving
two fundamentals) are expected to be stronger than third or higher order
combinations.  Indeed, the apparent absence of significant features above
3200cm$^{-1}$ is in good agreement with the expected highest frequency
of $\sim 1600$cm$^{-1}$ for the fundamental modes \cite{phystoday}.

Since anharmonic effects may, in principle, come into play due to the
excitations created by the IR radiation itself, variations in the spectra
as a function of the illumination intensity were checked for.  The gradual
decrease of the source intensity, typical of the synchrotron source, proved
to be very valuable in this respect.  The IR lines were found to be
independent of the intensity of the incident radiation, indicating that the
higher frequency modes are intrinsic to the sample.

The energy of a particular
combination mode $\nu_1\otimes \nu_2$ is generally expected to be $\approx
E(\nu_1) \pm E(\nu_2)$ (where $E(\nu)$ is the energy of mode $\nu$).
The temperature dependence of the line intensity of the combination modes
allows us to exclude the possibility that difference frequencies are seen in
the spectra.  The Bose factors for fundamentals at $\nu _1$ and $\nu _2$,
n$_1$ and n$_2$, respectively, appear in the intensity of the
addition mode at $E(\nu _1) + E(\nu _2)$ as $1+$n$_1 + $n$_2$, whereas
for the difference mode $E(\nu _1) - E(\nu _2)$ the intensity is
$\vert $n$_1 - $n$_2\vert $.  Consequently, for the wavenumber range
studied, the intensity of the difference modes should exhibit a strong
decrease with temperature.  None of the lines exhibit this behavior,
therefore we must conclude that all of the resonances seen in our
measurement are either fundamentals or should be close to the sum of
two fundamental frequencies.

The orientational ordering of the C$_{60}$ molecules is known to influence
the line shapes and intensities of the IR \cite{irrotation,kamaras1} and
Raman \cite{ramanrotation,kuzmany,kamaras1} modes.  Comparison of the
300K spectrum to the 77K spectrum suggests
that most of the low frequency lines exhibit dramatic narrowing at low
temperature, while the width of the typical high frequency line changes much
less.  Detailed analysis shows that in most of the cases the total
oscillator strength of the lines does not change, but the linewidth drops
sharply at around the orientational ordering transition temperature (Figure
\ref{77Kfit}).  This is in contrast to the gradual increase a peak heights
reported by Kamar\'as {\it et al.} \cite{kamaras1}.

In many cases the broader high temperature line splits to
several sharper low temperature resonances.  This is illustrated in Figure
\ref{tdep}, where spectra at several temperatures are plotted over an
expanded scale for some particularly interesting frequency ranges.  In the
two middle panels of Fig.
\ref{tdep} a few lines where the behavior is non-typical, in the sense that
the oscillator strength seems to increase dramatically below the phase
transition, are also pointed out.  These lines were either very
broad, or forbidden at high temperature.  The significance of these
resonances will be discussed later.

\subsection{Pressure Dependence}

Typical spectra obtained under pressure are displayed in Figure
\ref{pspectra}.  The $1100-1400$ and $1900-2250$cm$^{-1}$ regions are
omitted because the diamonds used in the pressure cell did not transmit
at those energies.  Most of the modes visible in the original
crystal spectra (Figure \ref{IR&modes}) are again seen under
pressure, however their energies shift some under pressure.
The three visible F$_{1u}$ modes' pressure dependencies were compared to
previously reports \cite{c60pres} and were
thus used to determine the approximate pressure being applied to the sample.
In Figure \ref{pdep} the center frequency positions of each of the major
modes observed in the pressure measurement are presented as a function of
applied pressure.  Most modes are observed to stiffen under pressure as is
usually the case.  The higher energy modes generally increase in energy
about twice as much as the lower energy modes, supporting our assumption
that they are binary combination modes.

There are a few new modes appearing under pressure that persist after the
pressure has been released.  These are illustrated in Figure \ref{newfromp}
by arrows.  Some are modes that were seen to be especially temperature
dependent as well (668, 764, and 1567cm$^{-1}$) indicating that the
rotational freezing and pressure could be having the same effect on
these modes.  Since the sample is no longer a single crystal and is
instead compressed to fill a gasket, it is very likely the internal
pressure is still larger than ambient making an accurate comparison
difficult.

The resonance at 611cm$^{-1}$ is totally absent in the single crystal
sample, and becomes very prominent after the pressure treatment.  No
corresponding low temperature line has been observed.  As we argue later,
lines that strongly increase in strength at low temperatures/high pressures
are probably {\it ungerade} fundamentals.  The pressure dependent line
611cm$^{-1}$ could be a candidate for a new mode, but there is no
corresponding evidence in the neutron spectrum \cite{neutron1}.
A combination frequency fits this mode well, but its strong pressure
dependence remains unexplained.

\section{Discussion and Mode Assignments}

The fact that modes are observed at exactly the frequencies where
Raman-active modes are known to be strongly suggests that IR-silent modes
are becoming weakly-IR-active.
These Raman modes must be becoming weakly active due to a
symmetry breaking.  The room temperature phase
has $T^{3}_{h}$ fcc structure \cite{t3} and the low temperature phase (below
$\sim 250$K) has either $T^6_h$ sc \cite{sc} or $T^4_h$ 2a$_0$-fcc
\cite{2a0fcc} structure.  All of these phases retain the inversion symmetry
on the molecular site which requires infrared and Raman spectroscopies be
strictly complimentary.  Thus the observation of Raman lines in our infrared
spectra indicates that they are not being activated by crystal field effects.
They are probably becoming active due to $^{13}$C isotopic substitution or
some type of inhomogeneity.  Since these types of symmetry breakings do not
preserve the inversion symmetry, they would make modes both IR- and
Raman-active.  The observation of the Raman modes becoming IR-active
implies that other silent modes could be activated as well.

As discussed earlier, the fact that modes up to roughly twice the highest
predicted frequency of the fundamental modes are observed implies second
order modes are within the spectra.
Group theory can be used to find out which
second order combinations are allowed to be IR-active.  The character tables
of the symmetry groups are used to determine the symmetry of second order
vibrational modes.  When a direct product of two fundamental modes contains
F$_{1u}$ symmetry in its character, it means that a combination of those two
modes is IR-active.  All the direct products can be straight-forwardly
calculated using the character table of the icosohedral group $I_{h}$
\cite{grouptheory}.  An example of a combination mode that has F$_{1u}$
character and thus is symmetry allowed is
\begin{equation}
F_{2g}\otimes H_u = {\bf F_{1u}}\oplus F_{2u} \oplus G_u \oplus H_u\\
\end{equation}
Similarly, the combinations that are Raman-active have A$_g$ or H$_g$ in
their character.  Table \ref{modesmix} summarizes which combination modes
are either IR-active (I) or Raman-active (R).

In the evaluation of the data we started with the 77K spectrum, since it has
narrow lines and more features than the spectra above the orientational
transition temperature.  In order to assign the weak lines to combination
modes and other silent modes becoming IR-active, we first selected the
values of the known fundamentals from other experiments. The low
temperature Raman frequencies used were reported in Refs.
\cite{ramanrotation,kuzmanyraman}, the F$_{1u}$ modes were taken from
the work of Homes {\it et al.} \cite{irrotation}.
(in the present work the F$_{1u}$ modes are so saturated, that no
resonance frequency can be deduced with sufficient accuracy).  In these
experiments, some of the lines were found to be split at low temperature;
accordingly, a split line was used in our evaluation.
Next, the group theoretical
results in Table \ref{modesmix} were used to look for allowed combinations.
We first identify quite a few absorptions as belonging to F$_{1u}\otimes
$H$_g$ and A$_g\otimes $F$_{1u}$ combinations since these fundamentals'
frequencies are experimentally well known.
We found that most, but not all of these combinations indeed appear in the
spectra, some prominently.  In many cases the known low temperature
splittings of these fundamentals could also be matched to splittings
observed in their combinations,
confirming that the energies of the combination modes do not shift
significantly from $E(\nu _1) + E(\nu _2)$.  For example, the split line
in the 77K spectrum at 2900cm$^{-1}$  corresponds to the the sum of the
(unsplit) A$_g(2)$ and the (split) F$_{1u}(4)$ frequencies, and becomes
a single line at room temperature \cite{caution}.
The remaining modes must now be matched up to silent vibrations and
combinations involving silent vibrations.

According to Table \ref{modesmix}, there are 380 allowed IR combinations
and 484 allowed Raman combination (not considering
the low temperature split of some resonances).  On the other hand, there are
only $\sim 100$ experimentally known Raman and $\sim 200$ IR lines.
Therefore one has to be careful when trying to match the experiment
to the theory; assuming randomly distributed fundamentals up to
1600cm$^{-1}$, the average spacing between combination lines would be
8cm$^{-1}$ in the IR and and a 6cm$^{-1}$ in Raman, with larger
spacing at the low and high ends, and smaller spacing in the middle of the
band.  Thus the low and high frequency ranges provide greater constraints
on the choice of fundamentals while the middle range should be used with
caution.  In order to make the best use of experimental evidence, we
established a set of rules as follows.\\
{\it i.)} Fundamentals were searched for to fit all of our data and the
Raman data of Dong {\it et al.} \cite{eklundraman} simultaneously.  The
fit was accepted only when {\it all} lines were explained.\\
{\it ii.)} An infrared mode was said to be fit when a fundamental or
combination was within a tolerance of 1.5cm$^{-1}$, and Raman within
2.5cm$^{-1}$ (although most fit better). The larger tolerance for the
Raman data was to accommodate any temperature dependence of the
line position since the Raman data were taken at 20K.\\
{\it iii.)} No fundamentals were considered if the neutron scattering
data excluded that frequency range.  \\
{\it iv.)} Modes that are known to split, or modes observed to split at low
temperatures were kept split in the fitting procedure. If a combinations
involving a split fundamental is a unique explanation for an experimental
line, then the split should disappear at room temperature. \\
{\it v.)} No third or higher order combinations, or difference frequencies
are considered.

When a mode shows an enhanced temperature dependence (as
illustrated in Figure \ref{tdep}, as well a few more modes from Figure
\ref{IR&modes}), it is a good candidate for a fundamental mode which is
being further activated by the rotational freezing ({\it i.e.} by crystal
field effects).  It would thus be an odd parity mode since it is observed
in the infrared.  Other candidates were modes that appeared in both IR and
Raman measurements at the same frequency, being activated by a symmetry
breaking that does not preserve inversion.  Finally two high frequency
modes were chosen to fit a neutron peak and several high energy
combination modes.  The parity of each mode was settled on by observing
how well it fit the IR and Raman data in the second order using either
parity.  When {\it every} IR and Raman mode was fit by this procedure,
46 good candidates for the vibrational modes of C$_{60}$ had been obtained.

Finally the specific symmetry groups for each mode were chosen based
on which other modes it successfully mixed with.  For example, Table
\ref{modesmix} reveals that the F$_{1u}$ vibrations only mix with A$_g$,
F$_{1g}$, and H$_g$ to make IR-active combinations.  Since the A$_g$ and
H$_g$ modes are well known, there will be only three new modes remaining
that mix uniquely with the F$_{1u}$'s in the IR.  These are then assigned to
be F$_{1g}$ modes.  Now, in addition to F$_{1u}$, these F$_{1g}$ modes mix
with 1 A$_u$ and 7 H$_u$ modes.  Continuing in this manner, we were able to
make probable assignments for all 46 modes.  The frequencies, assigned
symmetries, and which experiments measure each fundamental are listed in
Table \ref{fundamentals}.  Also shown are any low temperature splittings
used in the fits.  Nearly all the fundamentals were found do a very good
job fitting the higher order data, however this fit is probably not the
only possible one.

Table \ref{bigtable} presents all the experimentally observed modes from IR
and Raman and their assignments found as described above.
Many of the experimental lines have multiple assignments within tolerance.
When an obvious best fit was available it was placed in Table
\ref{bigtable}, however in a few cases two possible explanations for a
single experimental line were put in.  The many allowed combinations
that did not fit experimental values are not listed.  There are a few
very high energy modes that were not fit indicating the possibility that
some modes could be third (or higher) order combinations.

In recent measurements of Rb$_1$C$_{60}$ \cite{rb1c60}, we explored a
metal-insulator transition in a quenched phase.  When in this
insulating state, many new modes appear.  While the understanding
of this spectrum is still incomplete, it is likely that the quenched
phase has a more complex structure, with more than one fullerene in the
unit cell, resulting in a significant reduction of the symmetry.  Also,
in a system where potentially mobile electrons are present, weak modes may
be amplified by the charge transfer between fullerenes \cite{rice}.
These processes activate fundamentals, which become visible when the
metallic shielding is removed below the metal-insulator transition
temperature ($\sim 20$C).  The the ionization of the C$_{60}$, and the
presence of the alkaline metal ion may well shift some of the modes from the
original, pure fullerene values, as it was observed for the F$_{1u}(4)$ line,
or leave them unchanged, as seen for other IR and Raman lines.
Figure \ref{neutrondoping}, illustrates that the new modes compare well
with the neutron scattering data (taken from Ref. \cite{neutron1}) and with
the 46 modes in Table \ref{fundamentals}.

Figure \ref{neutrondoping} also demonstrates how our 46 modes are in
agreement with the neutron data. When comparing the IR and neutron results,
one has to consider that the neutron time-of-flight spectrometry integrates
over all wavenumbers, while the optical fundamentals are probed at
{\bf k}=0.  Some of the fundamentals derived in our
work are also in agreement with the high energy electron energy loss
spectroscopy (HREELS) data of Lucas {\it et al.} \cite{eels}. It appears
that the HREELS lines at 355, 444, 686, 758, 968, 1097, 1258, and
1565cm$^{-1}$ have corresponding resonances in our spectra at frequencies
shifted to lower values by 2-15cm$^{-1}$.

\section{Conclusions}

In summary, we have measured a very rich and complex IR spectrum of thick
C$_{60}$ crystals between 300 and 77K and under pressure up to $\sim
25$KBar.  We have assigned the 46 vibrational modes that agree with neutron
measurements.  We have used group theoretical predictions for second order
combination modes to assign {\it all} the observed (IR and Raman)
vibrational modes to specific symmetry fundamentals and combination modes.

In addition to the 14 IR- and Raman-active modes, many fundamentals have
been deduced from weak lines observed earlier in various IR and Raman
measurements.  As our analysis demonstrates, the combination modes at
difference frequencies are not visible in the spectra, and therefore it is
very reasonable to assume that most of the weak low frequency modes are in
fact fundamentals.  This observation explains the good agreement between
our results and the results obtained by Kamar\'as {\it et al.}
\cite{kamaras1}

In matching the combination modes to the experimental data we did not find
evidence for a significant shift from the ``ideal" value of
$E(\nu _1\otimes \nu_ 2)=E(\nu _1) + E(\nu _2)$.  Without a detailed
theory, it is impossible to guess how strong the anharmonicity is
since there is no simple relationship between the strength of a combination
mode and its frequency shift.

Some of the lines become extremely narrow at low temperatures.  We found,
in agreement with high resolution IR studies of the allowed resonances
\cite{irrotation}, that the crystal field splitting of the lines is about
ten wavenumbers at most.  We believe that lines separated by more than this
should be treated as separate modes or combinations. The widths of the
resonances above the rotational transition temperature seem to correlate
with the magnitude of the splitting of the lines below the transition,
indicating that crystal fields influence the high temperature linewidth.
For combination modes another source of broadening could be the
{\bf k} dependent dispersion of the contributing fundamentals (note that
fundamentals with $\pm${\bf k} contribute to {\bf k}=0 processes).

The knowledge of all fundamentals is crucial for testing model calculations
of fullerene vibrations.  By the study of weakly active fundamentals
one can also measure most of the Raman-active resonances
(including the oxygen and polymerization sensitive A$_g(2)$ mode)
without exposing the sample to a laser beam which could induce unwanted
changes in the sample.  In addition, this IR study of C$_{60}$ crystals
could be helpful in interpreting experiments on doped fullerenes where
many new resonances appear.  An exploration of the weakly active optical
modes in doped samples could eventually lead to a more complete
understanding of electron phonon coupling in these materials.

\acknowledgments

This work was supported by NSF grant DMR9202528.  We would like to thank
Philip B. Allen and K. Kamar\'as for valuable discussions.  Gwyn P. Williams'
and G. Larry Carr's assistance in running the spectrometers at the NSLS is
appreciated.  Thanks to Gene Dresselhaus and Peter Eklund for providing us
with their papers prior to publication.

\onecolumn
\narrowtext
\begin{figure}
\caption{Simulation of transmission and reflection measurements on a thick
sample.  A typical F$_{1u}$ vibration is simulated at 570cm$^{-1}$ and a
much weaker mode typical of the modes to be studied is simulated at
630cm$^{-1}$.  Plasma frequency ($\omega _p$) and width ($\Gamma $) for each
are labeled.}
\label{transrefl}
\end{figure}

\widetext
\begin{figure}
\caption{C$_{60}$ single crystal infrared transmission spectra at
0.5cm$^{-1}$ resolution obtained at 300K and 77K.}
\label{IR&modes}
\end{figure}

\narrowtext
\begin{figure}
\caption{IR transmission spectrum of a C$_{60}$ crystal at 77K (solid line)
and the fit to the data (dashed line).  Inset shows the temperature
dependence of the width $\Gamma $ of this mode illustrating the narrowing
occurring sharply below the rotational phase transition temperature
$\sim 250$K.}
\label{77Kfit}
\end{figure}

\widetext
\begin{figure}
\caption{Portions of the IR transmission spectra of a crystal at various
temperatures above and below the rotational phase transition temperature
$\sim 250$K.  The left panel illustrates the sharpening of many vibrational
lines below this temperature.  The two middle panels point out (with arrows)
two modes that seem to suddenly appear below this temperature.  The
rightmost panel presents the lack of temperature dependence of higher energy
modes; some show a general sharpening or splitting of modes at much
lower temperatures (77K) than the rotational phase transition.}
\label{tdep}
\end{figure}

\begin{figure}
\caption{IR transmission spectra of thick C$_{60}$ at $\sim 25$KBar and
after the pressure had been slowly released to ambient pressure.  The
omitted sections ($1100-1400$ and $1900-2250$cm$^{-1}$) are where the
diamonds did not transmit.}
\label{pspectra}
\end{figure}

\narrowtext
\begin{figure}
\caption{The pressure dependence of the center frequencies, $\omega _0$, of
the larger C$_{60}$ vibrational modes.  The three visible F$_{1u}$ modes are
labeled.}
\label{pdep}
\end{figure}

\begin{figure}
\caption{Top curves are C$_{60}$ crystal 300K spectra taken from
Figure 2.  Lower curves are detailed from Figure 5 showing the spectra
after high pressure was released to approximately ambient conditions.
Several new modes that have appeared after this procedure are pointed
out by arrows.}
\label{newfromp}
\end{figure}

\widetext
\begin{figure}
\caption{Comparison of the 46 modes from Table II to
neutron measurements (taken from Ref. [13]) 
and to an insulating, quenched Rb$_1$C$_{60}$ IR spectrum.  The known Raman
and IR modes are slightly offset vertically and labeled.  The modes in
the Rb$_1$C$_{60}$ IR spectrum with an asterix $(\ast )$ are known to
be the F$_{1u}(4)$ at higher doping levels and are {\it not} new
fundamentals.}
\label{neutrondoping}
\end{figure}

\narrowtext
\begin{table}
\caption{Vibrational mode species that mix to be IR- (I) or Raman- (R)
active in the second order.}
\begin{tabular}{c|cccccccccc}
&\multicolumn{5}{c}{Even Modes}&\multicolumn{5}{c}{Odd Modes}\\
&A$_g$&F$_{1g}$&F$_{2g}$&G$_g$&H$_g$&A$_u$&F$_{1u}$&F$_{2u}$&G$_u$&H$_u$\\
\tableline
A$_g$&R\\
F$_{1g}$&&R\\
F$_{2g}$&&R&R\\
G$_g$&&R&R&R\\
H$_g$&&R&R&R&R\\
A$_u$&&I&&&&R\\
F$_{1u}$&I&I&&&I&&R\\
F$_{2u}$&&&&I&I&&R&R\\
G$_u$&&&I&I&I&&R&R&R\\
H$_u$&&I&I&I&I&R&R&R&R&R\\
\end{tabular}
\label{modesmix}
\end{table}

\newpage
\begin{table}
\caption{The 46 vibrational modes of C$_{60}$ and their tentative
assignments used to fit the IR, Raman, and Neutron data.}
\begin{tabular}{ccc}
Energy (cm$^{-1}$)&Assignment&Experiments where Observed\\
\tableline
\multicolumn{3}{c}{Even Modes}\\
\tableline
495&A$_g(1)$&\\
1470&A$_g(2)$&\\
272 267&H$_g(1)$&\\
431&H$_g(2)$&\\
709&H$_g(3)$&\\
775 778&H$_g(4)$&\\
1102&H$_g(5)$&\\
1252&H$_g(6)$&\\
1425 1418&H$_g(7)$&\\
1576 1567&H$_g(8)$&\\
485&G$_g(1)$&IR, Raman, Neutron\\
541&F$_{2g}(1)$&IR, Neutron\\
568&F$_{1g}(1)$&IR, Raman, Neutron\\
764&F$_{2g}(2)$&IR, Raman, Neutron\\
961&G$_g(2)$&IR, Raman, (Neutron)\\
973&F$_{1g}(2)$&IR, Raman, Neutron\\
1199&G$_g(3)$&IR, Raman, Neutron\\
1214&F$_{2g}(3)$&IR, Neutron\\
1330&G$_g(4)$&IR, Neutron\\
1345&G$_g(5)$&IR, Raman, Neutron\\
1479 1484&F$_{1g}(3)$&IR, Raman, Neutron\\
1544&F$_{2g}(4)$&IR, Raman, Neutron\\
1596&G$_g(6)$&Neutron\\
\tableline
\multicolumn{3}{c}{Odd Modes}\\
\tableline
526&F$_{1u}(1)$&\\
577&F$_{1u}(2)$&\\
1183&F$_{1u}(3)$&\\
1429 1433&F$_{1u}(4)$&\\
342&H$_u(1)$&IR, Raman, Neutron\\
353&F$_{2u}(1)$&IR, Raman, Neutron\\
402&G$_u(1)$&IR, Raman, Neutron\\
579&H$_u(2)$&Raman, (Neutron)\\
664 668&H$_u(3)$&IR, Neutron\\
712&F$_{2u}(2)$&IR, Raman (Neutron)\\
739&G$_u(2)$&IR, Raman, Neutron\\
753&G$_u(3)$&IR, Neutron\\
797&F$_{2u}(3)$&IR, Raman, (Neutron)\\
828&H$_u(4)$&IR, Neutron\\
1038&F$_{2u}(4)$&IR, Neutron\\
1080&G$_u(4)$&IR, Raman, Neutron\\
1122&A$_u(1)$&IR, Neutron\\
1222&H$_u(5)$&IR, Neutron\\
1242&H$_u(6)$&IR, (Neutron)\\
1290&G$_u(5)$&IR, Raman, Neutron\\
1313&F$_{2u}(5)$&IR, Neutron\\
1526&G$_u(6)$&IR, Raman, Neutron\\
1600&H$_u(7)$&Neutron\\
\end{tabular}
\label{fundamentals}
\end{table}

\newpage
\widetext
\begin{table}
\caption{Table of all low temperature experimental modes and their group
theoretical assignments.}
\begin{tabular}{|ccc|ccc|ccc|ccc|}
\multicolumn{2}{c}{Data (cm$^{-1}$)}&Group Theory&
\multicolumn{2}{c}{Data (cm$^{-1}$)}&Group Theory&
\multicolumn{2}{c}{Data (cm$^{-1}$)}&Group Theory&
\multicolumn{2}{c}{Data (cm$^{-1}$)}&Group Theory\\
IR&Raman$^a$ &Assignment&IR&Raman$^a$ &Assignment&IR&Raman$^a$
&Assignment&IR&Raman$^a$ &Assignment\\
\tableline
& 273 &H$_g(1)$& 1242 &&H$_u(6)$& 1818 &&H$_g(7)\otimes $G$_u( 1)$&& 2331
&H$_g(8)\otimes $F$_{2g}( 2)$\\
 342 &&H$_u( 1)$&& 1251 &H$_g(6)$& 1830 &&H$_g(6)\otimes $F$_{1u}(2)$& 2335
 &&G$_g( 6)\otimes $G$_u( 2)$\\
& 343 &H$_u( 1)$& 1260 &&H$_g(2)\otimes $H$_u(4)$&& 1841 &H$_g(8)\otimes
$H$_g( 1)$& 2350 &&G$_g( 6)\otimes $G$_u( 3)$\\
 353 &&F$_{2u}( 1)$&& 1289 &G$_u( 5)$& 1843 &&F$_{2g}( 2)\otimes $G$_u(
 4)$&& 2350 &H$_g(8)\otimes $H$_g(4)$\\
& 355 &F$_{2u}( 1)$& 1290 &&G$_u( 5)$& 1854 &&H$_g(4)\otimes $G$_u( 4)$&
2368 &&G$_g( 4)\otimes $F$_{2u}( 4)$\\
 403 &&G$_u( 1)$& 1308 &&H$_g(1)\otimes $F$_{2u}( 4)$&& 1857 &H$_g(7)\otimes
 $H$_g(2)$& 2382 &&G$_g( 5)\otimes $F$_{2u}( 4)$\\
& 404 &G$_u( 1)$& 1310 &&H$_g(1)\otimes $F$_{2u}( 4)$&& 1875 &H$_g(5)\otimes
$H$_g(4)$& 2393 &&G$_g( 6)\otimes $F$_{2u}( 3)$\\
 431 &&H$_g(2)$&& 1310 &F$_{2u}( 5)$& 1876 &&F$_{2g}( 3)\otimes $H$_u( 3)$&&
 2393 &G$_u( 4)\otimes $F$_{2u}( 5)$\\
& 433 &H$_g(2)$& 1314 &&F$_{2u}(5)$& 1882 &&F$_{2g}( 3)\otimes $H$_u(3)$&
2409 &&G$_g( 4)\otimes $G$_u( 4)$\\
 485 &&G$_g( 1)$& 1314 &&G$_g( 2)\otimes $F$_{2u}( 1)$&  1884 &&F$_{2g}(
 3)\otimes $H$_u(3)$& 2422 &&G$_g( 3)\otimes $H$_u(5)$\\
& 486 &G$_g( 1)$& 1330 &&G$_g( 4)$& 1890 &&H$_g(3)\otimes $F$_{1u}( 3)$&
2433 &&H$_g(6)\otimes $F$_{1u}( 3)$\\
&&& 1342 &&F$_{2g}( 2)\otimes $H$_u( 2)$& 1893
 &&H$_g(3)\otimes $F$_{1u}( 3)$&& 2438 &G$_g( 2)\otimes $F$_{1g}( 3)$\\
& 497 &A$_g( 1)$& 1344 &&G$_g( 5)$& &&&  2462 &&H$_g(7)\otimes
$F$_{2u}( 4)$\\
 526 &&F$_{1u}( 1)$&& 1346 &H$_g(4)\otimes $F$_{1g}( 1)$&& 1901
 &H$_g(2)\otimes $A$_g( 2)$&& 2463 &H$_u(6)\otimes $H$_u(5)$\\
& 533 &H$_g( 1)\otimes $H$_g( 1)$& 1352 &&H$_g(1)\otimes $G$_u( 4)$&& 1913
&H$_g( 7)\otimes $A$_g( 1)$&& 2506 &H$_g( 6)\otimes $H$_g( 6)$\\
 542 &&F$_{2g}( 1)$& 1352 &&H$_g(4)\otimes $F$_{1u}( 2)$&  1915
 &&H$_g(6)\otimes $H$_u( 3)$&& 2527 &H$_g(7)\otimes $H$_g(5)$\\
 567 &&F$_{1g}( 1)$& & 1368 &H$_g(5)\otimes $H$_g( 1)$& 1925 &&A$_g(
 1)\otimes $F$_{1u}( 4)$& 2530 &&H$_g(5)\otimes $F$_{1u}( 4)$\\
& 570 &F$_{1g}( 1)$& 1376 &&H$_g(3)\otimes $H$_u(3)$& 1938 &&G$_g( 3)\otimes
$G$_u( 2)$&& 2551 &H$_g(8)\otimes $F$_{1g}( 2)$\\
 575 &&F$_{1u}( 2)$& &&& & 1959 &H$_g(6)\otimes $H$_g( 3)$&  2563
 &&H$_g(6)\otimes $F$_{2u}( 5)$\\
& 580 &H$_u( 2)$& 1394 &&F$_{1g}( 1)\otimes $H$_u(4)$&& 1959
&F$_{1u}(4)\otimes $F$_{1u}( 1)$&& 2570 &H$_g( 5)\otimes $A$_g( 2)$\\
 609 &&H$_g( 1)\otimes $H$_u( 1)$& &&&  1960 &&H$_g(4)\otimes $F$_{1u}(
 3)$& 2608 &&H$_g(7)\otimes $F$_{1u}( 3)$\\
 668 &&H$_u(3)$&& 1410 &H$_u(3)\otimes $G$_u( 2)$& 1968 &&H$_g( 8)\otimes
 $G$_u( 1)$&& 2611 &F$_{1u}( 4)\otimes $F$_{1u}( 3)$\\
& 692 &H$_u( 1)\otimes $F$_{2u}( 1)$& &&&  1979 &&H$_g(8)\otimes $G$_u(
1)$& 2624 &&F$_{2g}( 4)\otimes $G$_u( 4)$\\
 709 &&H$_g(3)$& 1418 &&H$_g(7)$& 1985 &&F$_{2g}( 2)\otimes $H$_u(5)$&& 2629
 &H$_g(7)\otimes $F$_{2g}( 3)$\\
& 711 &F$_{2u}( 2)$&& 1426 &H$_g(7)$& 1990 &&H$_g(6)\otimes $G$_u( 2)$& 2640
&&H$_g(7)\otimes $H$_u(5)$\\
 712 &&F$_{2u}( 2)$& 1429 &&F$_{1u}( 4)$& 1992 &&H$_g(6)\otimes $G$_u( 2)$&&
 2653 &H$_u(5)\otimes $F$_{1u}( 4)$\\
&&& & 1450 &G$_u( 2)\otimes $F$_{2u}( 2)$& 1999 &&H$_g( 3)\otimes
 $G$_u( 5)$& 2662 &&F$_{1g}( 3)\otimes $F$_{1u}( 3)$\\
&&&  1470 &&A$_g( 2)$&& 2006 &F$_{1u}( 4)\otimes $F$_{1u}( 2)$&& 2677
 &H$_g(7)\otimes $H$_g(6)$\\
 739 &&G$_u( 2)$&& 1470 &A$_g( 2)$& 2011 &&G$_g( 1)\otimes $G$_u( 6)$&& 2677
 &H$_g(8)\otimes $H$_g( 5)$\\
& 742 &H$_u( 1)\otimes $G$_u( 1)$& 1480 &&F$_{1g}( 3)$& 2016
&&H$_g(4)\otimes $H$_u(6)$& 2682 &&H$_g(6)\otimes $F$_{1u}( 4)$\\
 753 &&G$_u( 3)$&& 1481 &H$_g( 1)\otimes $F$_{2g}( 3)$&& 2025 &F$_{2u}(
 5)\otimes $F$_{2u}( 2)$& 2715 &&H$_g(7)\otimes $G$_u( 5)$\\
& 758 &H$_g(1)\otimes $G$_g( 1)$& 1484 &&F$_{1g}(3)$& 2028 &&G$_g( 3)\otimes
$H$_u(4)$&& 2717 &G$_u( 5)\otimes $F$_{1u}( 4)$\\
&&&  1497 &&F$_{1g}( 2)\otimes $F$_{1u}( 1)$&& 2041 &H$_g( 3)\otimes
 $G$_g( 4)$& 2730 &&H$_g(7)\otimes $F$_{2u}( 5)$\\
 764 &&F$_{2g}( 2)$&& 1502 &G$_g( 2)\otimes $F$_{2g}( 1)$& 2042 &&F$_{2g}(
 3)\otimes $H$_u(4)$&& 2736 &H$_g(6)\otimes $F$_{1g}(3)$\\
 775 &&H$_g(4)$& 1503 &&F$_{2g}( 2)\otimes $G$_u( 2)$& 2048 &&A$_g(
 2)\otimes $F$_{1u}( 2)$& 2740 &&F$_{2g}( 3)\otimes $G$_u( 6)$\\
& 775 &H$_g(4)$& 1509 &&H$_g( 1)\otimes $H$_u(6)$&& 2052 &G$_u( 6)\otimes
$F$_{1u}( 1)$&& 2773 &H$_g(8)\otimes $G$_g( 3)$\\
 796 &&F$_{2u}( 3)$& 1513 &&H$_g(1)\otimes $H$_u(6)$& 2063
 &&F$_{1g}(3)\otimes $H$_u( 2)$& 2778 &&H$_g(6)\otimes $G$_u( 6)$\\
& 798 &F$_{2u}( 3)$&& 1516 &F$_{2g}( 1)\otimes $F$_{1g}( 2)$&& 2070
&H$_u(6)\otimes $H$_u(4)$&& 2782 &H$_u(7)\otimes $F$_{1u}( 3)$\\
 827 &&H$_u(4)$&  1517 &&H$_g( 4)\otimes $G$_u( 2)$&& 2070 &H$_g(8)\otimes
 $A$_g( 1)$&  2790 &&H$_g(8)\otimes $H$_u(5)$\\
 827 &&G$_g( 1)\otimes $H$_u( 1)$& 1525 &&G$_u( 6)$& 2079 &&H$_g(6)\otimes
 $H$_u(4)$& 2814 &&F$_{2g}( 3)\otimes $H$_u(7)$\\
& 862 &H$_g(2)\otimes $H$_g(2)$& 1532 &&H$_g(4)\otimes $G$_u( 3)$& 2085
&&G$_g( 1)\otimes $H$_u(7)$&& 2823 &H$_u(7)\otimes $H$_u(5)$\\
 911 &&F$_{1g}( 1)\otimes $H$_u( 1)$&& 1532 &H$_g(5)\otimes $H$_g(2)$& 2092
 &&H$_g(8)\otimes $F$_{1u}( 1)$& 2839 &&G$_g( 6)\otimes $H$_u(6)$\\
& 919 &H$_u( 1)\otimes $F$_{1u}( 2)$& 1539 &&G$_g( 2)\otimes $H$_u( 2)$&
2099 &&G$_g( 5)\otimes $G$_u( 3)$&& 2850 &H$_g(7)\otimes $H$_g(7)$\\
 956 &&H$_g(2)\otimes $F$_{1u}( 1)$&& 1547 &H$_g( 4)\otimes $H$_g( 4)$&&
 2119 &G$_u( 4)\otimes $F$_{2u}( 4)$& 2856 &&H$_g(7)\otimes $F$_{1u}(4)$\\
 962 &&G$_g( 2)$& 1548 &&F$_{1g}( 2)\otimes $F$_{1u}( 2)$& 2123 &&F$_{2g}(
 4)\otimes $H$_u( 2)$& 2856 &&G$_g( 4)\otimes $G$_u( 6)$\\
& 962 &G$_g( 2)$& 1556 &&F$_{2g}( 3)\otimes $H$_u( 1)$&& 2136 &H$_g(
8)\otimes $F$_{1g}( 1)$& 2890 &&H$_g(8)\otimes $F$_{2u}( 5)$\\
 973 &&F$_{1g}( 2)$& 1563 &&H$_g(1)\otimes $G$_u( 5)$& 2137 &&H$_g(3)\otimes
 $F$_{1u}( 4)$&& 2896 &H$_g(7)\otimes $A$_g( 2)$\\
& 976 &H$_g(3)\otimes $H$_g( 1)$& 1567 &&H$_g(8)$& 2139 &&H$_g(3)\otimes
$F$_{1u}( 4)$& 2899 &&A$_g( 2)\otimes $F$_{1u}( 4)$\\
 1020 &&H$_g( 1)\otimes $G$_u( 3)$& 1572 &&H$_g(4)\otimes $F$_{2u}( 3)$&
 2151 &&H$_g(8)\otimes $F$_{1u}(2)$& 2901 &&A$_g( 2)\otimes $F$_{1u}(4)$\\
& 1022 &H$_u(3)\otimes $F$_{2u}( 1)$& 1576 &&H$_g(8)$&& 2167 &G$_u(
2)\otimes $F$_{1u}( 4)$& 2914 &&F$_{1g}(3)\otimes $F$_{1u}( 4)$\\
 1039 &&F$_{2u}( 4)$&& 1577 &H$_g(8)$& 2168 &&F$_{1g}( 1)\otimes $H$_u(7)$&
 && \\
& 1040 &F$_{2u}( 4)$& 1612 &&H$_g(2)\otimes $F$_{1u}( 3)$& 2172 &&H$_g(
7)\otimes $G$_u(3)$& 2930 &&G$_g( 4)\otimes $H$_u(7)$\\
 1061 &&H$_g(3)\otimes $F$_{2u}( 1)$&& 1619 &H$_g( 1)\otimes $G$_g( 5)$&
 2176 &&G$_g( 6)\otimes $H$_u( 2)$&& 2940 &A$_g( 2)\otimes $A$_g( 2)$\\
& 1068 &H$_u( 3)\otimes $G$_u( 1)$& 1628 &&H$_g(5)\otimes $F$_{1u}( 1)$&
2178 &&H$_g(7)\otimes $G$_u( 3)$& 2951 &&H$_g(7)\otimes $G$_u( 6)$\\
 1080 &&G$_u( 4)$&& 1634 &H$_u( 1)\otimes $G$_u( 5)$& 2180 &&H$_g(5)\otimes
 $G$_u( 4)$& 2994 &&H$_g(8)\otimes $F$_{1u}( 4)$\\
& 1080 &G$_u( 4)$& &&& & 2180 &H$_g(3)\otimes $A$_g( 2)$&& 3002
&H$_g(8)\otimes $H$_g(7)$\\
\end{tabular}
\label{bigtable}
\end{table}

\begin{table}
\caption{Continuation of Table III.}
\begin{tabular}{|ccc|ccc|ccc|ccc|}
\multicolumn{2}{c}{Data (cm$^{-1}$)}&Group Theory&
\multicolumn{2}{c}{Data (cm$^{-1}$)}&Group Theory&
\multicolumn{2}{c}{Data (cm$^{-1}$)}&Group Theory&
\multicolumn{2}{c}{Data (cm$^{-1}$)}&Group Theory\\
IR&Raman\tablenote{Raman data taken from reference \cite{eklundraman} }
&Assignment&IR&Raman$^a$ &Assignment&IR&Raman$^a$ &Assignment&IR&Raman$^a$
&Assignment\\
\tableline
1100 &&H$_g(5)$& &&& &&&  3004 &&H$_g(8)\otimes $F$_{1u}( 4)$\\
& 1101 &H$_g(5)$& 1671 &&G$_g( 4)\otimes $H$_u( 1)$& 2195 &&F$_{1g}(
2)\otimes $H$_u(5)$&& 3046 &H$_g(8)\otimes $F$_{1g}( 3)$\\
 1115 &&H$_g(4)\otimes $H$_u( 1)$& 1680 &&H$_g(5)\otimes $F$_{1u}( 2)$&&
 2199 &H$_g(7)\otimes $H$_g(4)$&& 3046 &H$_g(8)\otimes $A$_g(2)$\\
 1121 &&A$_u( 1)$&& 1693 &H$_g( 1)\otimes $H$_g( 7)$& 2205 &&H$_g(4)\otimes
 $F$_{1u}( 4)$&& 3118 &H$_g(8)\otimes $F$_{2g}( 4)$\\
& 1141 &G$_u( 2)\otimes $G$_u( 1)$& 1694 &&H$_g( 1)\otimes $F$_{1u}( 4)$&
2213 &&F$_{2g}( 4)\otimes $H$_u(3)$& 3128 && \\
 1142 &&H$_g(2)\otimes $F$_{2u}( 2)$& 1713 &&G$_g( 2)\otimes $G$_u( 3)$&&
 2227 &F$_{2u}( 3)\otimes $F$_{1u}( 4)$&& 3134 &H$_g(8)\otimes $H$_g(8)$\\
 1150 &&G$_g( 1)\otimes $H$_u( 3)$& 1720 &&H$_g(2)\otimes $G$_u( 5)$& 2235
 &&H$_g(3)\otimes $G$_u( 6)$& 3140 && \\
 1153 &&G$_g( 1)\otimes $H$_u(3)$& 1728 &&G$_g( 1)\otimes $H$_u(6)$& 2240
 &&H$_g(8)\otimes $H$_u( 3)$&& 3152 &H$_g(8)\otimes $H$_g(8)$\\
 1168 &&H$_g(2)\otimes $G$_u( 2)$&& 1736 &H$_g(1)\otimes $A$_g( 2)$&& 2243
 &F$_{2g}( 2)\otimes $F$_{1g}( 3)$& 3241 && \\
 1175 &&H$_g(4)\otimes $G$_u( 1)$& 1747 &&H$_g( 3)\otimes $F$_{2u}( 4)$&
 2245 &&H$_g(8)\otimes $H$_u(3)$& 3265 && \\
 1182 &&F$_{1u}( 3)$& 1750 &&F$_{1g}( 1)\otimes $F$_{1u}( 3)$& 2263 &&G$_g(
 6)\otimes $H$_u(3)$&& 3267 & \\
& 1187 &H$_u( 3)\otimes $F$_{1u}( 1)$& 1760 &&H$_g(7)\otimes $H$_u( 1)$&&
2274 &H$_g(8)\otimes $H$_g(3)$& 3280 && \\
 1199 &&G$_g( 3)$&& 1776 &H$_g(2)\otimes $G$_g( 5)$& 2275 &&G$_g( 2)\otimes
 $F$_{2u}( 5)$&& 3282 & \\
& 1204 &H$_g(3)\otimes $A$_g( 1)$& 1782 &&F$_{2g}( 1)\otimes $H$_u(6)$& 2280
&&H$_g( 8)\otimes $F$_{2u}( 2)$&& 3325 & \\
 1205 &&F$_{2g}( 1)\otimes $H$_u( 3)$&& 1789 &H$_u( 3)\otimes $A$_u( 1)$&&
 2286 &H$_g(8)\otimes $H$_g(3)$& 3350 && \\
 1214 &&F$_{2g}( 3)$& 1792 &&H$_g( 1)\otimes $G$_u( 6)$& 2293 &&F$_{2g}(
 3)\otimes $G$_u( 4)$&& 3385 & \\
 1223 &&H$_u(5)$& 1808 &&F$_{1g}( 1)\otimes $H$_u(6)$& 2311
 &&F$_{1g}(3)\otimes $H$_u(4)$& 3402 && \\
 1237 &&F$_{1g}( 1)\otimes $H$_u(3)$&& 1811 &H$_g( 1)\otimes $F$_{2g}( 4)$&
 2328 &&H$_g(8)\otimes $G$_u( 3)$&3442 && \\
\end{tabular}
\label{bigtable2}
\end{table}


\begin{references}
\bibitem{caution}As it happens the 2900cm$^{-1}$ line is uniquely assigned to
the F$_{1u}(4)$ - A$_g(2)$ combination, and therefore well suited for
this test.  Many times, especially in the middle of the frequency band, the
experimental lines have several closely matching combinations, making
the temperature dependence of line splittings harder to match uniquely.
\end{references}
\end{document}